\documentclass[conference]{IEEEtran}

\usepackage[export]{adjustbox}
\usepackage[dvipsnames]{xcolor}
\usepackage{empheq}
\usepackage{amsmath,amsfonts,amsthm,amssymb,amsbsy,paralist,ifthen}
\usepackage{mathrsfs}
\usepackage[mathcal]{euscript}
\usepackage{graphicx}
\usepackage{psfrag}
\usepackage{subfigure}
\usepackage{url}
\usepackage{stfloats}
\usepackage{float}
\usepackage{array}
\usepackage{algorithm} 
\usepackage{algorithmic} 
\usepackage{color}
\usepackage{cases}
\usepackage{bm}
\usepackage{colortbl}
\usepackage{arydshln}
\usepackage{multirow}
\usepackage{multicol}
\usepackage{cite}
\usepackage{extarrows}
\usepackage[table]{xcolor}

\begin{document}

\title{Physics-Informed Deep Learning for Channel Prediction in Satellite-Mounted Metasurface Reflection Communication Networks}

\author{Aolin Sun$^*$, Ruihong Jiang$^*$, Jincong Mo$^*$, Yangbo Guo$^\dag$ and Zhuangzhuang Cui$^\ddag$ \\

\small
$^*$State Key Laboratory of Networking and Switching Technology and Beijing Key Laboratory of Convergent \\ Communications and Networking Technologies in Low-Orbit Constellation Systems \\ Beijing University of Posts and Telecommunications, Beijing 100876, China \\
$^\dag$School of Computer and Information Engineering, Henan University, Kaifeng 475001, China\\
$^\ddag$Department of Electrical Engineering (ESAT), KU Leuven, Belgium.  
}


\maketitle

\begin{abstract}
Satellite-mounted metasurfaces (MS) provide a low-power programmable paradigm for 6G satellite-terrestrial networks, effectively compensating for severe free-space path loss under stringent payload constraints. However, accurate channel prediction for such systems remains fundamentally challenging due to the coupling of dynamic orbital topology, stochastic user geometry, and hardware-imposed phase quantization. This paper proposes a physics-informed deep learning (PIDL) framework that embeds the physics of wireless communication, including monotonicity constraints and asymptotic decay properties, into a deep neural network for outage probability (OP)-aware channel prediction. We construct a realistic system model incorporating spherical user distribution, Spherical Poisson Point Process-based satellite constellation, and practical metasurface quantization. Simulation results demonstrate that the proposed framework achieves over 0.99 prediction accuracy with physically consistent outage trends under varying geometric configurations, constellation densities, and quantization resolutions.

\end{abstract}

\begin{IEEEkeywords}
Satellite communications, metasurface, deep learning, performance prediction, stochastic geometry.
\end{IEEEkeywords}

\IEEEpeerreviewmaketitle

\section{Introduction}
\label{sec:introduction}

Satellite communication has emerged as a pivotal technology for achieving global seamless connectivity, particularly in remote and underserved regions where terrestrial infrastructure is unavailable or economically unfeasible \cite{maral2009satellite}. With the rapid development of low Earth orbit (LEO) mega-constellations, satellite-terrestrial integrated networks represent a cornerstone of sixth-generation (6G) wireless systems, promising global seamless coverage and enhanced service capacity \cite{zhu2017non}. However, the inherent characteristics of satellite-terrestrial links pose fundamental challenges that restrict further improvement of system performance, including significant path loss, limited onboard power, and dynamic channel conditions. Traditional solutions, such as higher transmit power or active on-board relays, face tight constraints in terms of satellite payload power budget, hardware cost and thermal dissipation capacity, necessitating innovative solutions to improve spectral and energy efficiency.

To address these challenges, reconfigurable metasurfaces (MS) have recently garnered substantial attention as a transformative technology for space reflection communications \cite{wu2019intelligent}. By intelligently manipulating the electromagnetic properties of incident signals, programmable MS can create favorable propagation environments without requiring active radio-frequency chains. This inherent feature brings prominent advantages, including ultra-low power consumption, low hardware complexity and flexible deployment, which are highly compatible with the resource-constrained satellite platform. Moreover, satellite-terrestrial links are dominated by line-of-sight (LoS) propagation, which enables MS to effectively compensate for long-distance path loss and enhance link robustness through coherent superposition of reflected signals via precise phase tuning \cite{wang2021reconfigurable}. Hence, satellite-mounted MS provides a promising paradigm for improving the coverage and reliability of space reflection communication systems.

While the potential of MS-assisted satellite communications has been recognized, previous theoretical investigations have primarily relied on idealized assumptions that limit their applicability to practical scenarios. In \cite{wang2024outage}, the authors derived closed-form outage probability (OP) expressions for static scenarios using stochastic geometry. Furthermore, stochastic user distribution modeling was incorporated in \cite{wang2025performance} to analyze the outage performance in more realistic coverage scenarios. These analytical frameworks, while mathematically rigorous, typically assume continuous phase control with infinite precision, fixed satellite positions, and simplified user distribution models. The transition from theoretical models to practical deployment introduces multiple non-linear factors that challenge traditional analytical approaches: (i) dynamic satellite topology with stochastic spatial distributions, (ii) localized user distribution with angular constraints in realistic coverage areas, and (iii) hardware-imposed discrete phase quantization.

The limitations of pure analytical methods in capturing these coupled complexities motivate the exploration of learning-based approaches for channel modeling and performance prediction. Physics-informed deep learning (PIDL), with its powerful ability to embed physical constraints into data-driven models, presents a compelling alternative for performance prediction in complex wireless systems \cite{oshea2017introduction}. By integrating domain knowledge of wireless communications with neural network approximation, PIDL ensures both high accuracy and physical interpretability \cite{yang2019deep}. 
In this context, \cite{corra2025low} proposes a physics-informed reinforcement learning method that embeds known system dynamics via post-decision states and stochastic sampling to reduce complexity. Meanwhile, \cite{huang2026physics} develops a physics-informed implicit neural representation for RIS-aided ISAC wireless imaging, integrating explicit physical forward models with channel-state-information-driven constraints.






Motivated by the above, we propose a PIDL framework for channel prediction in satellite-mounted MS reflection communications with practical constraints. The OP is employed as the key validation metric to assess prediction fidelity. The main contributions are summarized as follows:
\begin{itemize}
    \item We develop a satellite-mounted MS space reflection model jointly incorporating Spherical Poisson Point Process (SPPP)-based dynamic satellite topology, angular-constrained spherical user distribution, and discrete MS phase quantization with limited precision.
    \item We construct a comprehensive dataset over a multidimensional feature space spanning transmit power, geometry, constellation scale, and hardware precision, and design a physics-informed neural network embedding wireless communication physics, including monotonicity and asymptotic decay constraints, for channel prediction validated via OP.
    \item We validate the proposed PIDL framework through numerical simulations under diverse geometric configurations, constellation densities, and phase quantization resolutions, demonstrating over 0.99 prediction accuracy with physically consistent outage trends.
\end{itemize}


\section{System Model}
\label{sec:system}

\begin{figure}[!t]
\centering
\includegraphics[width=0.75\columnwidth]{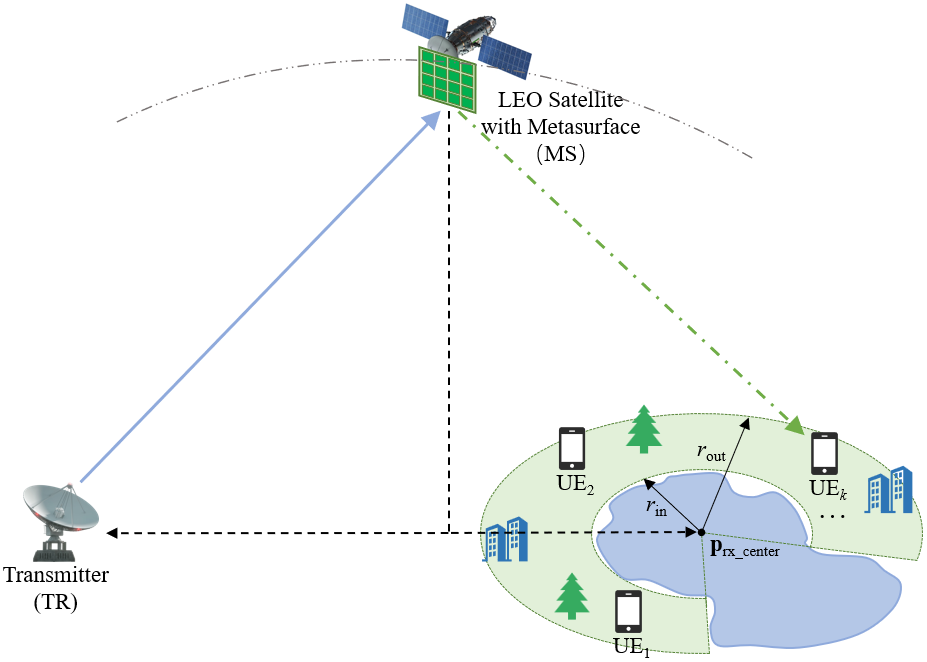}
\caption{Considered system model.}
\label{fig:systemmodel}
\end{figure}

We consider a downlink space reflection communication system, as shown in Fig.~\ref{fig:systemmodel}, where a terrestrial transmitter (TR) communicates with $K$ users (UEs) via a LEO satellite-mounted MS consisting of $L$ reflecting elements. Owing to 
potential non-line-of-sight blockages, direct ground-to-user transmission is infeasible. Conventional active relaying is limited by stringent satellite payload constraints on power, hardware complexity, and thermal dissipation. Instead, the MS acts as a programmable passive reflector without active radio-frequency chains, thereby circumventing these limitations. By judiciously tuning the phase shifts of its $L$ elements, the MS coherently superimposes reflected signal components to synthesize directional beamforming gain toward the intended receivers, establishing reliable reflection links with negligible power overhead.

To manage multi-user interference and channel state information uncertainty, the transmitter adopts rate-splitting multiple access (RSMA), whereby each user's message is decomposed into a common stream and a private stream \cite{lyu2024rate}. The MS phase configuration is then adapted to align the reflected signal components constructively at each intended receiver, which enhances the received signal-to-noise ratio and improves link reliability across the coverage area.

\subsection{Channel Model}
\label{subsec:channel}

The considered space reflection system features a two-hop cascaded channel (i.e., TR-MS-UE links), accounting for both large-scale path loss and small-scale fading. 

For large-scale propagation, the path loss for the TR-MS or MS-UE link is modeled as
\begin{flalign}
\mathcal{L}\left({{d_i}} \right) = G_i \left(\frac{c}{4\pi f_\textrm{c}}\right)^2 d_i^{-\vartheta},
\end{flalign}
where $i \in \{\mathrm{t},\mathrm{r}\}$; $d_\mathrm{t}$ and $d_\mathrm{r}$ represent the distances of TR-MS and MS-UE links, respectively; $G_\mathrm{t}$ and $G_\mathrm{r}$ are the antenna gain of the transmitter and the receiver, respectively; $c$ is the speed of light, $f_\textrm{c}$ is the carrier frequency, and $\vartheta$ is the path loss exponent.

For small-scale fading, the equivalent cascaded channel gain for the $k$-th UE under ideal phase alignment is given by
\begin{flalign}
|\tilde{h}|^2 = \left(\sum\nolimits_{l=1}^{L}|h_l||g_{l,k}|\right)^2,
\end{flalign}
where $h_l$ and $g_{l,k}$ represent the TR-MS and MS-UE channel coefficients, respectively, modeled as independent shadowed-Rician random variables.

\subsection{Transmission Model}
\label{subsec:rsma}


Based on the cascaded channel characterized above, the transmitter adopts RSMA to serve $K$ users. Specifically, each user's message is split into a common stream $s_\mathrm c$ and a private stream $s_{\textrm{p},k}$, with $\mathbb{E}[|s_{\textrm{c}}|^2]=\mathbb{E}[|s_{\textrm{p},k}|^2]=1$\cite{lyu2024rate}. The transmitted signal is
\begin{flalign}
\mathbf{x}=\sqrt{P_\mathrm c}s_\mathrm c+\sum\nolimits_{k=1}^{K}\sqrt{P_{\textrm{p},k}}s_{\textrm{p},k},
\end{flalign}
where $P_\mathrm c$ and $P_{\textrm{p},k}$ are the transmit power allocated to the common stream and the $k$-th private stream, respectively.

Let $P_\mathrm{t}$ denote the total transmit power and $\alpha_c$ denote the power allocation factor for the common stream. Then, the common-stream power is $P_\mathrm c=\alpha_\mathrm c P_\mathrm t$, and the private-stream power is $P_{\textrm{p},k}=\frac{(1-\alpha_\textrm{c})P_\mathrm{t}}{K}$ for all $k$ with equal power allocation. The received signal at the $k$-th UE is given by
\begin{flalign}
y_k = \sqrt{ \mathcal{L}(d_{\rm t}) \mathcal{L}(d_{\rm{r}}) g_\mathrm{s}} |\tilde{h}|^2  \mathbf{x} + N_0,
\end{flalign}
where $g_\mathrm{s}$ is the spread spectrum gain, $N_0$ is the additive white Gaussian noise (AWGN) with variance $\sigma^2$. The received signal-to-interference-plus-noise ratios (SINR) for the common and private streams at the $k$-th UE are given by
\begin{flalign}
\gamma_{\textrm{c},k} = \frac{|\tilde{h}|^2\rho P_\textrm{c}}{\sum_{i=1}^{K}|\tilde{h}|^2\rho P_{\textrm{p},i}+\sigma^2},
\label{eq:SINR_common}
\end{flalign}
and
\begin{flalign}
\gamma_{\textrm{p},k} = \frac{|\tilde{h}|^2\rho P_{\textrm{p},k}}{\sum_{i\neq k}^{K}|\tilde{h}|^2\rho P_{\textrm{p},i}+\sigma^2},
\label{eq:SINR_private}
\end{flalign}
respectively, where $\rho = \mathcal{L}(d_\mathrm{t})\mathcal{L}(d_{\mathrm{r}})g_\mathrm{s}$ represents the composite link gain factor incorporating path loss and spreading gain. Accordingly, the achievable rates at the $k$-th UE are given by $C_{\rm{c},k} = W \log_2(1+ \gamma_{\textrm{c},k})$ for the common stream and $C_{\rm{p},k} = W \log_2(1+ \gamma_{\textrm{p},k})$ for its private stream, with $W$ being the available bandwidth of the system.

\subsection{Outage Probability Analysis}
\label{subsec:outage}

Given the achievable rates derived above, the OP is defined as the probability that the instantaneous information rate falls below the target thresholds for either the common or private stream, i.e.,
\begin{flalign}
P_\mathrm{out} = 1 - \Pr\{C_{\textrm{c},k} \geq R_{\textrm{c},k}, C_{\textrm{p},k} \geq R_{\textrm{p},k}\},
\end{flalign}
where $R_{\textrm{c},k}$ and $R_{\textrm{p},k}$ denote the target thresholds for the common and private streams, respectively.


By using moment matching, the cascaded channel gain can be approximated by a Gamma distribution \cite{le2025secrecy}, enabling the derivation of closed-form outage expressions \cite{wang2025performance}. The high-SNR asymptotic analysis yields that the diversity gain is determined by $G_\mathrm{d} = \alpha L / 2$ \cite{wang2025performance}, where $\alpha$ denotes the shape parameter of the approximate Gamma distribution. While these closed-form results provide valuable asymptotic insights and scaling laws, they rely on several idealized assumptions that limit their applicability to real-world scenarios: (i) continuous phase control with infinite precision, (ii) fixed satellite positions, and (iii) simplified circular user distribution. The following section addresses these practical constraints through a data-driven methodology.

\section{System Model with Practical Constraints}
\label{sec:practical}

\begin{figure}[!t]
\centering
\includegraphics[width=0.55\columnwidth]{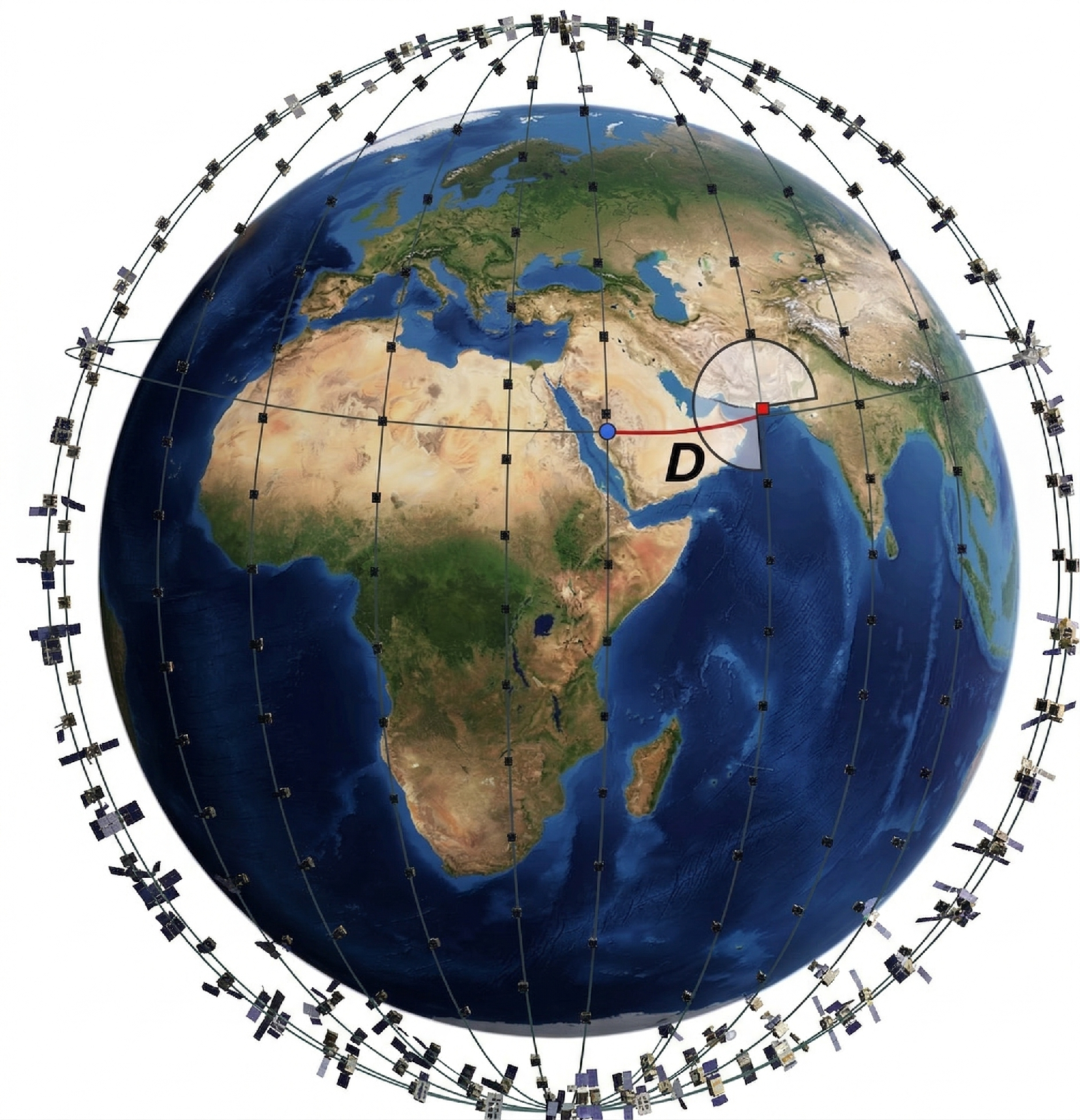}
\caption{Considered system model with practical constraints.}
\label{fig:sys_PIDL}
\end{figure}

This section extends the theoretical framework in Section \ref{sec:system} to incorporate three critical practical constraints: dynamic satellite topology, realistic user distribution, and hardware quantization limitations. The specific system model is illustrated in Fig.~\ref{fig:sys_PIDL}.

\subsection{Spherical Poisson Point Process for Satellite Topology}
\label{subsec:sppp}
To capture the stochastic spatial distribution of mega-constellations such as Starlink, we model the satellite topology using a Spherical Poisson Point Process (SPPP) on a spherical shell $\mathcal{S}_{\mathrm{sat}}$ of radius $R_{\mathrm{s}} = R_{\mathrm{E}} + H$, where $R_{\mathrm{E}}$ denotes Earth's radius and $H$ is the orbital altitude. For any measurable subset $\mathcal{A} \subseteq \mathcal{S}_{\mathrm{sat}}$, the number of satellites $N(\mathcal{A})$ follows a Poisson distribution with intensity measure $\Lambda(\mathcal{A}) = \int_{\mathcal{A}} \lambda_{\mathrm{s}}(\mathbf{s}) \, d\mathbf{s}$, i.e.,
\begin{equation}
\mathbb{P}\{N(\mathcal{A}) = n\} = \frac{(\Lambda(\mathcal{A}))^n}{n!} e^{-\Lambda(\mathcal{A})},
\label{eq:sppp}
\end{equation}
where $\lambda_s(\mathbf{s})$ denotes the spatial density function.

Given a constellation of size $N_{\mathrm{s}}$, the uniform distribution density on the spherical surface becomes a constant, i.e., $\lambda = N_{\mathrm{s}} / [4\pi R_{\mathrm{s}}^2]$, which reflects the average number of satellites per unit area on the spherical surface. This naturally accounts for Earth's curvature and enables accurate slant range and elevation angle calculations critical to wide-area path loss modeling \cite{kim2026spectrum}. In each realization, ground receivers associate with the nearest visible satellite based on geometric distance, thereby introducing randomness in serving node selection that fundamentally departs from fixed-topology analysis.






\subsection{Ring-Sector User Distribution Model}
\label{subsec:user}


To capture such spatial heterogeneity, we model UEs as uniformly distributed in a ring-sector region $\mathcal{R}$ centered at the receiver center point $\mathbf{p}_\text{rx\_center}$. The region is geometrically parameterized by inner radius $r_\text{in}$, outer radius $r_\text{out}$, start angle $\psi_0$, and angular span $\Delta\psi$, enabling flexible characterization of both coverage range and directional service patterns.

To ensure uniform areal density on the spherical Earth surface and eliminate edge-region sampling bias, the radial distance $s$ is generated via the square-root transformation, i.e.,
\begin{equation}
s = \sqrt{u \cdot (r_\text{out}^2 - r_\text{in}^2) + r_\text{in}^2}, \quad u \sim \mathcal{U}[0,1].
\label{eq:radial_sampling}
\end{equation}
This transformation preserves constant probability density per unit surface area, ensuring statistical fairness across the coverage region. Subsequently, the azimuth angle $\psi$ is drawn uniformly from $[\psi_0, \psi_0 + \Delta\psi]$, and the tangent plane direction is determined by the local orthogonal basis $\{\mathbf{e}_1, \mathbf{e}_2\}$ at the center point:
\begin{equation}
\mathbf{t}_\mathrm{dir} = \cos(\psi)\, \mathbf{e}_1 + \sin(\psi)\, \mathbf{e}_2.
\end{equation}

The user position is then mapped onto the spherical Earth surface to enforce strict geometric consistency:
\begin{equation}
\mathbf{p} = \cos\!\left(\frac{s}{R_\textrm{E}}\right) \mathbf{p}_\text{rx\_center} + R_\textrm{E} \sin\!\left(\frac{s}{R_\textrm{E}}\right) \mathbf{t}_\mathrm{dir},
\label{eq:spherical_mapping}
\end{equation}
followed by radial normalization $\mathbf{p}_{\text{user}} = R_\textrm{E} \cdot \mathbf{p}/\|\mathbf{p}\|$ to correct numerical deviations and guarantee that all generated coordinates lie precisely on the Earth's surface.

\subsection{MS Phase Quantization Model}
\label{subsec:quantization}

In addition to dynamic topology and realistic user distribution, practical MS implementations face hardware constraints on phase resolution \cite{ramezani2025joint}. Let $B$ denote the quantization bit budget. The achievable phase set is $\mathcal{F} = \{0, \Delta\theta, \ldots, (2^B-1)\Delta\theta\}$ with the step size $\Delta\theta = 2\pi/2^B$.

The hardware maps ideal continuous phases to the nearest quantization levels via
\begin{flalign}
\phi_q = \text{round}\left(\frac{\phi_\text{ideal}}{\Delta\theta}\right) \cdot \Delta\theta,
\label{eq:quantization}
\end{flalign}
introducing random phase error $\Delta\phi = \phi_\text{ideal} - \phi_q$. This discretization causes vector misalignment during signal superposition, resulting in non-linear power gain degradation:
\begin{flalign}
\left\|\sum\nolimits_{l=1}^{L} g_{l,k} h_l e^{j\phi_{q,l}}\right\|^2.
\end{flalign}
The quantization error becomes particularly severe at low bit budgets (1-2 bits), while higher precision (3-4 bits) approaches ideal continuous performance.

\section{Proposed Physics-Informed Learning Framework}
\label{sec:method}



\begin{figure}[!t]
\centering
\includegraphics[width=0.80\columnwidth]{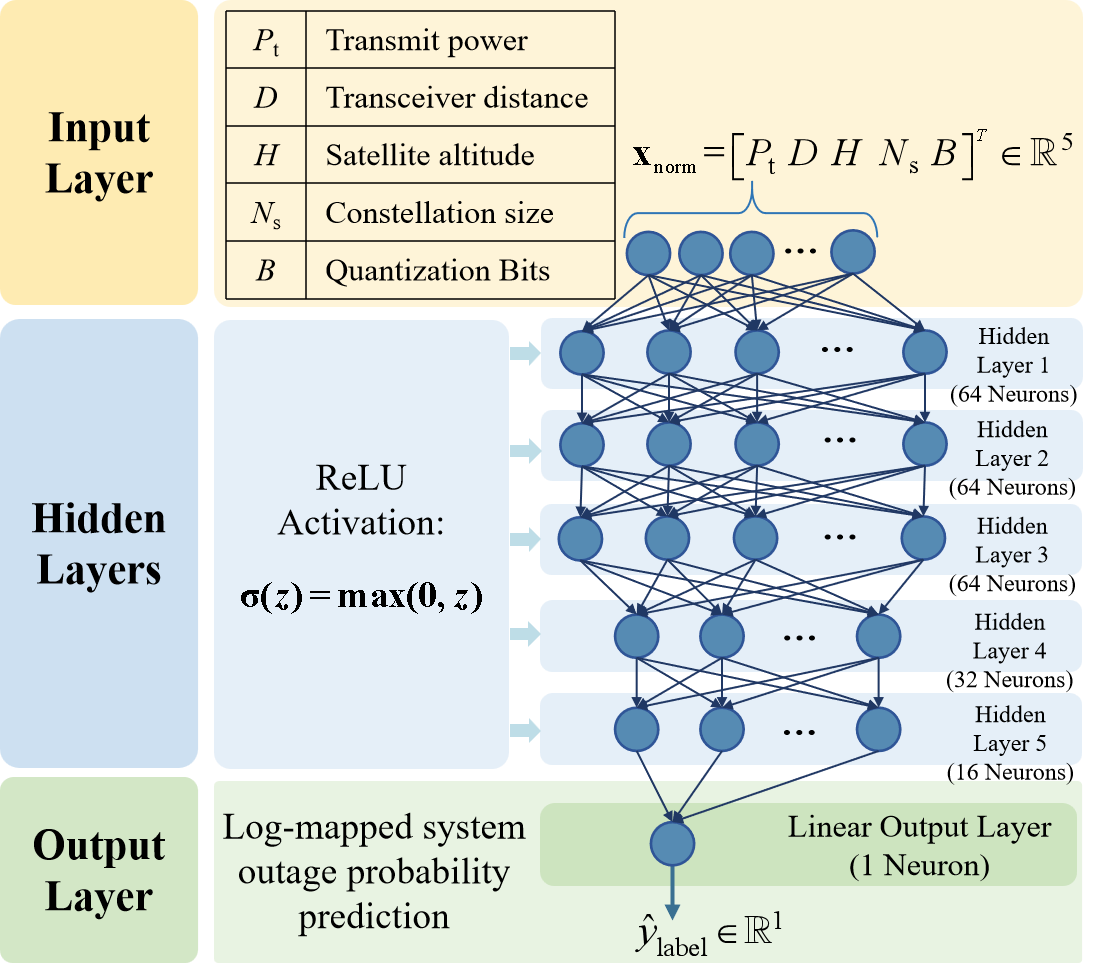}
\caption{Proposed PIDL architecture for OP-aware channel prediction.}
\label{fig:network}
\end{figure}

To overcome the analytical intractability arising from the coupled practical constraints, we propose a PIDL framework for OP-aware channel prediction. As illustrated in Fig.~\ref{fig:network}, the framework integrates data-driven neural network approximation with domain-specific physical constraints, ensuring both high prediction accuracy and physical consistency. The overall pipeline consists of four stages: dataset generation, feature preprocessing, neural network architecture design, and physics-informed loss optimization.

\subsection{Dataset Generation} \label{subsec:dataset}

We generate a comprehensive dataset by sampling the multi-dimensional parameter space through large-scale Monte Carlo simulations. The input feature vector is defined as
\begin{equation}
\mathbf{x} = [P_\mathrm{t}, D, H, N_\mathrm{s}, B]^{\mathsf{T}} \in \mathbb{R}^{5},
\end{equation}
where each dimension corresponds to a key physical parameter with the following ranges:
\begin{itemize}
    \item Transmit power: $P_\mathrm{t} \in [45, 60]$ dBm, capturing the power budget variation,
    \item Transceiver distance: $D \in [300, 500]$ km, reflecting the ground link geometry,
    \item Satellite altitude: $H \in [300, 500]$ km, accounting for constellation topology dynamics,
    \item Constellation size: $N_\mathrm{s} \in [2000, 10000]$, characterizing stochastic satellite availability,
    \item Quantization resolution: $B \in \{1, 2, 3, 4\}$ bits, modeling practical MS hardware constraints.
\end{itemize}

The dataset comprises 9,600 samples, partitioned into training (70\%), validation (20\%), and test (10\%) sets. To ensure statistical reliability of the OP labels, each sample is averaged over $10^{6}$ independent Monte Carlo channel realizations.

\subsection{Feature Preprocessing} \label{subsec:preprocessing}
\textit{Feature Normalization:} Due to disparate physical scales across the input parameters (e.g., power in dBm versus constellation size in thousands), we apply min-max normalization to map all features to the $[0,1]$ interval:
\begin{equation}
x_{\text{norm}} = \frac{x - x_{\min}}{x_{\max} - x_{\min}}.
\label{eq:minmax}
\end{equation}
This standardization ensures stable gradient updates and prevents dominant features from biasing the learning process.

\textit{Logarithmic Label Transformation:}
The OP $P_{\text{out}}$ spans an extremely wide dynamic range ($10^{0}$ to $10^{-100}$). Direct regression on raw probability values would suffer from numerical instability and poor sensitivity in deep outage regions. To address this, we apply a logarithmic transformation:
\begin{equation}
y_{\text{label}} = \log_{10}(P_{\text{out}} + \delta),
\label{eq:log_transform}
\end{equation}
where $\delta = 10^{-30}$ is a small offset preventing numerical overflow at $P_{\text{out}} \approx 0$. This transformation converts exponential probability decay into a linear regression target, significantly improving prediction fidelity across the entire dynamic range.

\subsection{Neural Network Architecture} \label{subsec:architecture}
We design a five-layer fully connected deep neural network with a tapered architecture $\{64, 64, 64, 32, 16\}$, as depicted in Fig.~\ref{fig:network}. The network topology is structured as follows:

\begin{itemize}
    \item \textbf{Input Layer:} A 5-dimensional normalized feature vector $\mathbf{x}_{\text{norm}} \in \mathbb{R}^{5}$.
    \item \textbf{Hidden Layers:} Five fully connected layers with ReLU activation $\sigma(z) = \max(0,z)$, which mitigates the vanishing gradient problem and enables efficient training of deep architectures.
    \item \textbf{Output Layer:} A single linear neuron producing the predicted log-outage $\hat{y}_{\text{label}} \in \mathbb{R}^{1}$.
\end{itemize}

The layer-wise forward propagation is given by
\begin{equation}
\mathbf{h}^{(l)} = \sigma\!\left(\mathbf{W}^{(l)}\mathbf{h}^{(l-1)} + \mathbf{b}^{(l)}\right), \quad l = 1, \ldots, 5,
\label{eq:forward}
\end{equation}
where $\mathbf{W}^{(l)}$ and $\mathbf{b}^{(l)}$ denote the weight matrix and bias vector of the $l$-th layer, respectively, with $\mathbf{h}^{(0)} = \mathbf{x}_{\text{norm}}$.

\subsection{Physics-Informed Loss Function and Training Strategy}
To ensure both high prediction accuracy and physical consistency, we embed wireless communication physics into the learning process. The total loss function is formulated as a weighted sum of three terms:
\begin{equation}
    \mathcal{L}_{\text{total}} = \mathcal{L}_{\text{Huber}} + \lambda_m\mathcal{L}_{\text{mono}} + \lambda_a\mathcal{L}_{\text{asym}},
\end{equation}
where $\lambda_m = 0.05$ and $\lambda_a = 0.02$ are hyperparameters balancing physical constraints.

For robust regression across the wide dynamic range of outage probabilities, the data-driven term employs the Huber loss (Smooth L1 Loss):
\begin{equation}
    \mathcal{L}_{\text{Huber}}(y, \hat{y}) = 
    \begin{cases}
        \frac{1}{2}(y - \hat{y})^2, & |y - \hat{y}| \leq \beta, \\
        \beta|y - \hat{y}| - \frac{1}{2}\beta^2, & \text{otherwise},
    \end{cases}
\end{equation}
where $\beta = 0.1$.

The monotonicity constraint enforces the physical property that OP decreases with transmit power $P_t$ and increases with distance $D$ and altitude $H$:
\begin{flalign}
     \mathcal{L}_{\text{mono}} =& \tfrac{1}{N_b}\sum\nolimits_{i=1}^{N_b}\big[\text{ReLU}\big(\tfrac{\partial \hat{y}^{(i)}}{\partial P_{t,\text{norm}}^{(i)}}\big)\big]^2  \\
     & + \big[\text{ReLU}\big(-\tfrac{\partial \hat{y}^{(i)}}{\partial D_{\text{norm}}^{(i)}}\big)\big]^2 + \big[\text{ReLU}\big(-\tfrac{\partial \hat{y}^{(i)}}{\partial H_{\text{norm}}^{(i)}}\big)\big]^2,   \notag
\end{flalign}
where gradients are computed via automatic differentiation.

The asymptotic decay constraint aligns the predicted high-SNR slope with the theoretical diversity order $G_d = \alpha L/2$ derived in Section II-C:
\begin{equation}
    \mathcal{L}_{\text{asym}} = \tfrac{1}{N_b}\sum_{i=1}^{N_b}\left|\tfrac{\log_{10}\hat{P}_{\text{out}}^{(i)}(P_{t,2}) - \log_{10}\hat{P}_{\text{out}}^{(i)}(P_{t,1})}{\log_{10}P_{t,2} - \log_{10}P_{t,1}} + \tfrac{G_d}{10}\right|^2,
\end{equation}
where $P_{t,2} > P_{t,1}$ are paired samples from the same batch.

The training configuration follows: Adam optimizer with ReduceLROnPlateau scheduler (decay factor 0.5 after 10 stagnant epochs), early stopping after 20 epochs without validation improvement, implemented in PyTorch.

\begin{table}[t!] 
\renewcommand{\arraystretch}{1.2}
\centering  
\caption{Simulation Parameters} 
 \begin{tabular}{l|c|c}  
     \hline
     \hline
      Parameters & Symbol & Values \\ 
       \hline
       Transmitter antenna gain & $G_\mathrm{t}$ & 32 dBi \\
        \cdashline{1-3}[0.8pt/2pt]
       Receiver antenna gain & $G_\mathrm{r}$& 15 dBi \\
       \cdashline{1-3}[0.8pt/2pt]
       Spreading gain & $g_\mathrm{s}$ & 30 dB \\
       \cdashline{1-3}[0.8pt/2pt]
       Carrier frequency & $f_\mathrm{c}$ & 1 GHz \\
       \cdashline{1-3}[0.8pt/2pt]
        Path loss exponent & $\vartheta$ & 2 \\
       \cdashline{1-3}[0.8pt/2pt]
       System bandwidth & $W$ & 20 MHz \\
       \cdashline{1-3}[0.8pt/2pt]
       Number of users & $K$ & 2 \\
       \cdashline{1-3}[0.8pt/2pt]
       Number of MS elements & $L$ & 100 \\
       \cdashline{1-3}[0.8pt/2pt]
       \multirow{2}{5cm}{The target rates of the common and private parts for the $k$-th UE} & $R_{\rm{c},k}$ & 30 kbps \\
       \cdashline{2-3}[0.8pt/2pt]
        & $R_{\rm{p},k}$ & 10 kbps \\
        \cdashline{1-3}[0.8pt/2pt]
       \multirow{3}{5cm}{The parameters of the shadowed-Rician fading channel} & $a$ & 0.251  \\
       \cdashline{2-3}[0.8pt/2pt]
        & $\Omega$ & 0.279 \\
       \cdashline{2-3}[0.8pt/2pt]
        & $m$ & 5 \\
       \cdashline{1-3}[0.8pt/2pt]
       \hline
       \hline
   \end{tabular}
   \label{tab:parameters}
\end{table}

\section{Simulation Results and Analysis}
\label{sec:simulation}

This section validates the proposed PIDL framework and investigates the impact of practical constraints on system performance. Unless otherwise specified, the default simulation parameters are listed in Table~\ref{tab:parameters}.


\begin{figure}[!t]
\centering
\includegraphics[width=0.75\columnwidth]{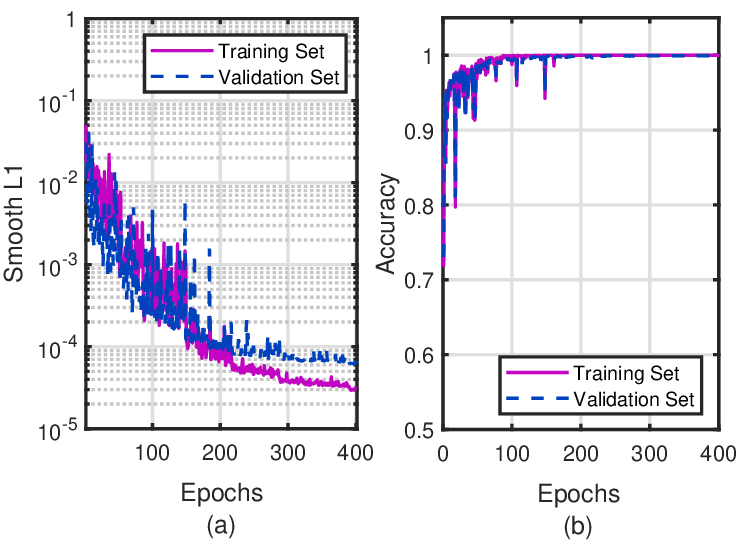}
\caption{PIDL training performance: (a) Huber loss, (b) prediction accuracy.}
\label{fig:convergence}
\end{figure}

Fig.~\ref{fig:convergence} illustrates the convergence behavior and training performance of the proposed PIDL model. As shown in the left subplot, the Smooth L1 Loss decreases monotonically for both the training and validation sets, rapidly converging to the order of $10^{-5}$ within 400 epochs. The close alignment between the two curves indicates that the model generalizes well without overfitting during training. Furthermore, the right subplot shows prediction accuracy, defined as the proportion of predicted log-transformed outage probabilities consistent with the Monte Carlo simulation labels, which rises sharply during the initial training phase and eventually stabilizes above 0.99. This indicates that the model predictions achieve a well-fitting metric against ground-truth outage labels, demonstrating the robust learning capability of our proposed model.



\begin{figure}[!t]
\centering
\includegraphics[width=0.75\columnwidth]{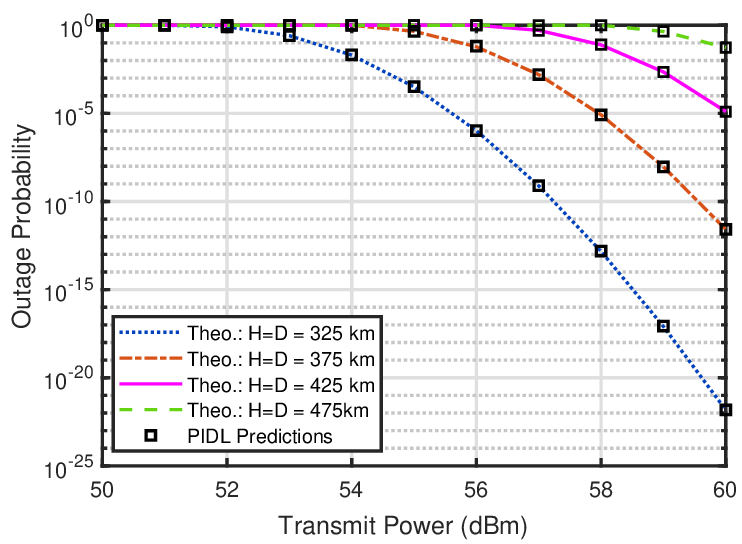}
\caption{OP vs. transmit power under varying geometric constraints.}
\label{fig:topology}
\end{figure}

Fig.~\ref{fig:topology} compares the PIDL predictions against the theoretical curves for varying satellite altitude $H$ and transceiver distance $D$, where $H=D=\{325,375,425,475\}$ km. It is seen that the OP decreases with transmit power but increases for larger $H$ and $D$ due to severe path loss. Crucially, the PIDL predictions align perfectly with theoretical baselines across all scenarios. Moreover, the model maintains high fidelity even in extreme outage regions ($10^{-20}$), confirming the effectiveness of logarithmic label preprocessing for wide-range regression.

\begin{figure}[!t]
\centering
\includegraphics[width=0.75\columnwidth]{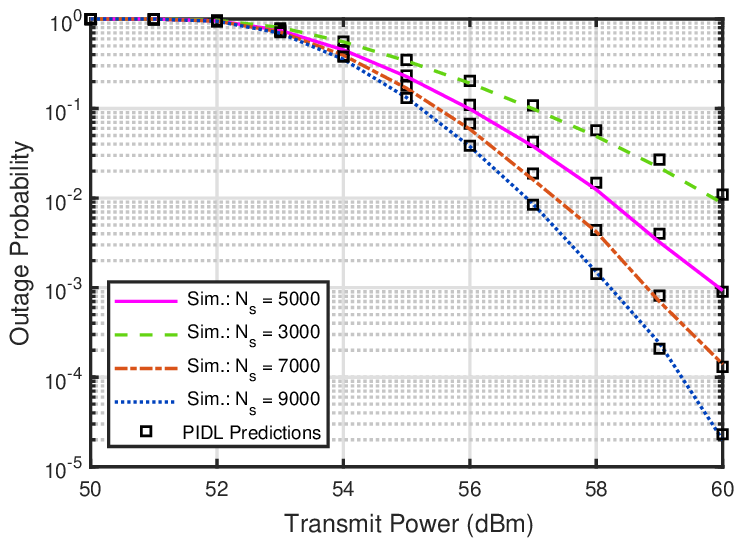}
\caption{Impact of satellite distribution intensity on outage performance.}
\label{fig:constellation}
\end{figure}

Fig.~\ref{fig:constellation} shows the impact of satellite distribution intensity on the OP performance. It is observed that as $N_\mathrm{s}$ rises from 3000 to 9000, the OP curves shift left significantly, as higher density in the SPPP topology allows UEs to find closer optimal nodes, reducing large-scale fading. Besides, the PIDL predictions align perfectly with simulations, confirming the model's ability to capture topological variations instantly. However, compared to ideal theoretical scenarios, performance remains lower even at $N_\mathrm{s} = 9000$. This highlights the inherent gap between simplified theoretical models and realistic random distributions, where satellites cannot guarantee optimal positioning.


\begin{figure}[!t]
\centering
\includegraphics[width=0.75\columnwidth]{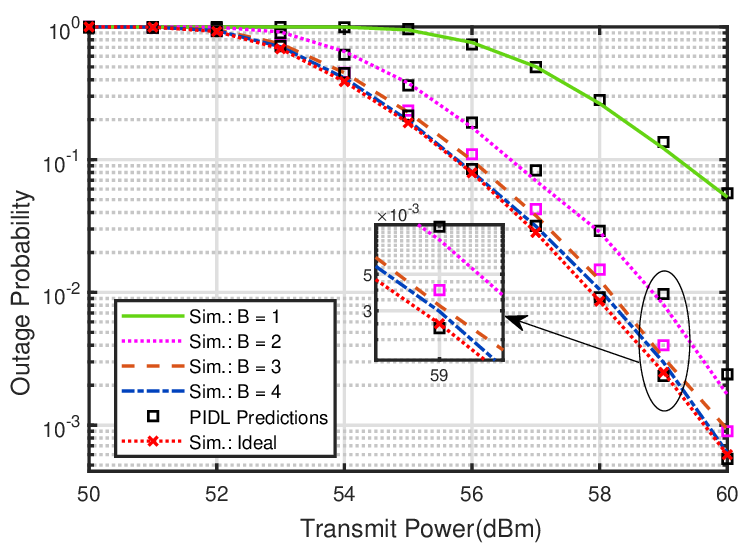}
\caption{Impact of MS phase quantization bits on outage performance.}
\label{fig:quantization}
\end{figure}

Fig.~\ref{fig:quantization} depicts the performance loss caused by MS phase quantization errors. It is seen that the 1-bit quantization scenario suffers significant power loss from severe phase misalignment compared to the ideal case. As $B$ increases, the phase error $\Delta\phi$ diminishes, and the system performance rapidly converges toward the ideal baseline. Meanwhile, the performance gap becomes negligible when $B \geq 3$, suggesting that 3 or 4 bits are sufficient for practical engineering implementations. Additionally, the PIDL predictions accurately fit these nonlinear performance transitions, further validating the model's precision in capturing quantization effects.


\section{Conclusion} \label{sec:conclusion}
This work presents a PIDL framework for channel prediction in satellite-mounted MS space reflection communication systems under practical constraints. By embedding stochastic satellite-user geometry and phase quantization into a neural network with physics-informed loss, the framework achieves over 0.99 prediction accuracy with physically consistent outage trends under varying geometric configurations, constellation densities, and quantization resolutions. This framework supports interpretable channel prediction and real-time optimization for 6G satellite-terrestrial integrated networks.


\section*{ACKNOWLEDGMENT}
This work is supported by the National Natural Science Foundation of China (NSFC) under Grant Nos. 62301077 and 62401201, and is also supported by the Research Foundation–Flanders (FWO), Senior Postdoctoral Fellowship under Grant No. 12AFN26N.

\bibliographystyle{IEEEtran}

\end{document}